\documentclass[12pt]{article}
\usepackage{curves}
\usepackage{amsmath, amssymb, lamsarrow}
\usepackage{amssymb}
\usepackage{tikz}
\usepackage{hyperref}
\def\mineappendix{
        \setcounter{section}{1}
        \setcounter{subsection}{0}
        \def\thesection{\Alph{section}}
        \def\sectionap{\@startsection  {section}{1}{\z@}
                        {-3.5ex plus-1ex minus-.2ex} {0ex plus.2ex}
                        {\reset@font\Large\bf  Appendix:  \, }
                        }
        }
\makeatother
\def\Proclaim #1. #2\par{\bigbreak\noindent{\sc#1.\enspace}{\it#2}\par}

\newtheorem{lem}{Lemma}[section]
\newtheorem{cor}[lem]{Corollary}
\newtheorem{thm}[lem]{Theorem}
\newtheorem{pro}[lem]{Proposition}
\newtheorem{exa}[lem]{Example}

\newtheorem{rem}[lem]{Remark}

\newcommand{\la}{\lambda}

\newenvironment{ack}{\noindent \textbf{Acknowledgments}.}

\title{Action of Virasoro operators on Hall-Littlewood polynomials}

\author{Xiaobo Liu \thanks{Research was partially supported by NSFC grants 11890662 and 11890660.}, Chenglang Yang}

\date{}

\begin{document}
\maketitle

\begin{abstract}
In this paper, we prove formulas for the action of Virasoro operators on Hall-Littlewood polynomials at roots of unity.
\end{abstract}


\section{Introduction}
\label{sec:intro}

Hall-Littlewood functions played important roles in representation theory of symmetric groups and finite general linear groups (c.f. survey article \cite{Mor76}). This class of symmetric functions includes ordinary Schur functions and Schur Q-functions as specializations. It is well known that these two special classes of functions give polynomial tau-functions for KP and BKP hierarchies (c.f. \cite{DJM} and \cite{Y}). Schur Q-functions were also used to study Kontsevich-Witten and Brezin-Gross-Witten tau functions which are generating functions of certain intersection numbers on
moduli spaces of stable curves (c.f. \cite{MM20}, \cite{Alex20}, \cite{Alex21}, \cite{LY}, \cite{LY2}).
Recently Mironov and Morozov proposed to use Hall-Littlewood functions specialized at roots of unity to study
generalized Kontsevich matrix models (c.f. \cite{MM21}).  An interesting problem is to
investigate whether Hall-Littlewood functions can also be used to study Gromov-Witten invariants since they are natural
generalizations of intersection numbers on moduli spaces of stable curves.
Virasoro constraints are powerful tools in the study of matrix models and Gromov-Witten invariants (c.f. \cite{EHX} and \cite{CK}). In fact
Kontsevich-Witten and Brezin-Gross-Witten tau functions are determined by the Virasoro constraints up to a scalar. This was the starting point for the proof of Q-polynomial expansion formulas for these two tau functions in the approach given in \cite{LY} and \cite{LY2}. To adapt this approach to more general models in Gromov-Witten theory and matrix models,
it is important to know how Virasoro operators acting on Hall-Littlewood functions. The main purpose of this paper is to give
formulas for the action of Virasoro operators on Hall-Littlewood functions specialized at roots of unity.

In this paper, we will consider Hall-Littlewood functions $Q_\la( \mathbf{t}; \rho)$ as polynomials of variables $\mathbf{t}=(t_1,t_2,...)$, where $r \, t_r$ is the
$r$-th power sum function in the theory of symmetric functions (c.f. MacDonald's book \cite{Mac}).
These polynomials are indexed by partitions $\la$ and also depend on a parameter $\rho\in\mathbb{C}$ (Here $\rho$ corresponds to the parameter $t$ in \cite{Mac}). They are Schur polynomials when $\rho=0$ and Schur Q-polynomials when $\rho=-1$.
If $\rho$ is the $n$-th root of unity $\xi_n$, we will denote the corresponding Hall-Littlewood polynomials by $Q^{(n)}_\la$, i.e.
\[ Q^{(n)}_\la(\mathbf{t})= Q_\la( \mathbf{t}; \xi_n).\]
When computing derivatives of $Q_\la$, it is convenient to extend the definition of $Q_\la$ to the case where $\la\in\mathbb{Z}^l$.
This was already considered in the original article \cite{L}. It is also rather natural from the view point of Jing's vertex operator approach to Hall-Littlewood polynomials in \cite{J}.

For any $n\geq2$ and $m \in \mathbb{Z}$, define
	\begin{align}
	L_m^{(n)}
    &:= \sum_{k \geq 1, \atop n\nmid k}kt_k\frac{\partial}{\partial t_{k+nm}}
	+\frac{1}{2}\sum_{k=1,\atop n\nmid k}^{mn-1}\frac{\partial^2}{\partial t_k \partial t_{mn-k}}
    - \frac{1}{2}\sum_{k=1, \atop n\nmid k}^{-mn-1} k(mn+k)t_k t_{-mn-k}  
	+\delta_{m,0}\frac{n^2-1}{24}. \label{eqn:Vira op xi_n}
	\end{align}
Here we set $t_k=0$ and $\frac{\partial}{\partial t_{k}} = 0$ if $k \leq 0$. Hence $L_m^{(n)}$ does not contain
quadratic terms $t_k t_{-mn-k}$ if $m \geq 0$, and it does not contain second order derivative terms if $m \leq 0$.
These operators form a Virasoro algebra since they satisfy the following bracket relation
\begin{align} \label{eqn:Vir-Bracket}
[L_i^{(n)},L_j^{(n)}]=n(i-j)L_{i+j}^{(n)}+\delta_{i+j,0}\frac{n^2(n-1)(i^3-i)}{12}
\end{align}
for all $i, j \in \mathbb{Z}$.
Note that operators $\frac{1}{n} L_m^{(n)}, m \in \mathbb{Z}, $ satisfy the standard Virasoro bracket relation.

 The first main result of this paper is  the following:
\begin{thm}\label{thm:Vira xi_n}
	For any $n \geq 2$,  $m \geq 0$, and $\la\in\mathbb{Z}^l$, we have
	\begin{align}\label{eqn:Vira xi_n}
	L_m^{(n)}  Q^{(n)}_\la
    =\sum_{i=1}^l \la_i Q^{(n)}_{\la-mn\epsilon_i}
	+ \sum_{k=1}^{mn-1}\sum_{i,j=1\atop i>j}^{l} (1-\xi_n^{-k})Q^{(n)}_{\la-k\epsilon_i-(mn-k)\epsilon_j}
    +\delta_{m,0}\frac{n^2-1}{24} Q^{(n)}_\la,
	\end{align}
	where
    \begin{equation} \label{eqn:la-a}
        \la- a \epsilon_i := (\la_1, \cdots, \la_i-a, \cdots, \la_l)
    \end{equation}
    for     $\la = (\la_1, \cdots,  \la_l)$ and $a \in \mathbb{Z}$.
\end{thm}

Since Virasoro constraints in Gromov-Witten theory and in matrix models usually start with the
$L_{-1}$-constraint which corresponds to the so called string equation,
it is also important to know the formula for $L_m^{(n)} Q^{(n)}_\la$ with $m<0$.
Let
$\mathcal{P}$ be the set of partitions, i.e. $(\la_1, \cdots, \la_l) \in \mathcal{P}$ if $\la_1 \geq \la_2 \geq \cdots \geq \la_l >0$.
The second main result of this paper is the following:
\begin{thm}\label{thm:-Vira xi_n}
For any $n\geq2$, $m \geq 1$, and $\la\in\mathbb{Z}^l$, we have
\begin{align}\label{eqn:-Vira xi_n}
L_{-m}^{(n)}  Q^{(n)}_\la
=&\sum_{i=1}^l \bigg(\la_i+\frac{m(n-1)}{2} \bigg) Q^{(n)}_{\la+mn\epsilon_i}
+\sum_{k=1, \atop n\nmid k}^{mn-1}\sum_{i,j=1\atop i>j}^{l} \xi_n^k Q^{(n)}_{\la+k\epsilon_i+(mn-k)\epsilon_j} \nonumber\\
&+\sum_{i=1}^l \sum_{k=1, \atop n\nmid k}^{mn-1} \xi_n^{k} \sum_{\mu \in \mathcal{P} \atop |\mu|=k} c_\mu(\xi_n) Q^{(n)}_{(\la+(mn-k)\epsilon_i,\mu)} \nonumber\\
&+\frac{1}{2}\sum_{k=1,\atop n\nmid k}^{mn-1}\Big(\sum_{\mu \in \mathcal{P} \atop |\mu|=k}\sum_{j=1}^{l(\mu)} c_{\mu}(\xi_n) Q^{(n)}_{(\la,\mu+(mn-k)\epsilon_j)}
+\sum_{\mu,\nu \in \mathcal{P} \atop |\mu|=k,|\nu|=mn-k} c_{\mu}(\xi_n) c_{\nu}(\xi_n) Q^{(n)}_{(\la,\mu,\nu)}\Big),
\end{align}
where $|\mu|:=\sum_{i} \mu_i$ if $\mu=(\mu_1, \mu_2, \cdots)$, and coefficients $c_\mu(\xi_n)$ will be given in Proposition~\ref{pro:multipr} (In fact, $c_\mu(\xi_n)$ are related to special cases of
Green's polynomials).
\end{thm}

Note that there are many ways to construct Virasoro operators. Our choice of Virasoro operators $L_{m}^{(n)}$ agrees with those used in
\cite{MM21}, where only $m>0$ cases were considered. These are essential parts of the operators used in the Virasoro constraints for
generalized Kontsevich matrix models. Other types of Virasoro operators might be obtained from $L_{m}^{(n)}$ by adding finitely many
pure derivative terms of order 1 or 2 for $m>0$. The action of such extra terms on Hall-Littlewood polynomials can be easily computed
from the well known derivative formula (see equation \eqref{eqn:derivative} below). For $m<0$, one may need to add finitely many
quadratic terms like $t_k t_m$ to obtain new Virasoro operators. The action of such quadratic terms on Hall-Littlewood polynomials
$Q_\la(\textbf{t}; \rho)$ can be easily computed from the following formula
\begin{align}\label{eqn:multi rtr}
p_r Q_{\la}=\sum_{i=1}^{l(\la)}Q_{\la+r\epsilon_i}
+\sum_{\mu \in \mathcal{P} \atop|\mu|=r}c_\mu(\rho)Q_{(\la,\mu)}
\end{align}
for all $\la\in\mathbb{Z}^l$,
where $p_r=r t_r$ corresponds to the $r$-th power sum function in the theory of symmetric functions, and
an explicit formula for coefficients $c_\mu(\rho)$ will be given in Proposition~\ref{pro:multipr}.
We will give a proof of this formula in section \ref{sec:pre} since we
could not find it in the literature.
For $\rho=0$, i.e. for the Schur polynomials, this formula is well known and is related to
Murnaghan-Nakayama rule  (c.f. \cite{Mac}). For $\rho=-1$, i.e. for Schur Q polynomials, this formula should  also be well known (see, for example,
\cite{B} and \cite{LY}). In Section 5 of \cite{B}, it was attempted but without success to find such a formula for general $Q_\la(\textbf{t}; \rho)$.

\begin{rem}
We notice that some formulas for $L_{m}^{(n)}  Q^{(n)}_\la$ with $m \geq 1$ were given in \cite{MM21} without proof under the assumption that $\la$ and $\la-km\epsilon_i - (n-k) m \epsilon_j$ for all $1 \leq k \leq n-1$ and $1 \leq i, j \leq l$
are strict partitions, i.e. components of these vectors are strictly decreasing and non-negative. Equation (50) in \cite{MM21} coincide with the corresponding formula in Theorem \ref{thm:Vira xi_n} for $m=1$ with the understanding that the order of components of $\la$ should be reversed in \cite{MM21}. For $m \geq 2$, it appears that some terms are missing in equations (51) and (52) in \cite{MM21}.
\end{rem}

 For $n=2$,  $Q^{(n)}_\la$ is the Schur Q polynomial. Formulas for the action of a different set of Virasoro operators
 on  $Q^{(2)}_\la$ have been given in \cite{ASY} and \cite{ASY2}. Note that variable $t_k$ in those papers corresponds to
 $2 t_k$ in the current paper. After this adjustment, the difference between $L_m^{(2)}$ and the corresponding operator
 in \cite{ASY} is $\frac{1}{4} \sum_{k=1}^{2m-1} \frac{\partial^2}{\partial t_k \partial t_{2m-k}}$ for $m > 0$.
 Using equation \eqref{eqn:derivative} below, it is easy to check that
 the main result in \cite{ASY} is equivalent to equation \eqref{eqn:Vira xi_n} for $n=2$.
 One can also use equation \eqref{eqn:multi rtr} to check the equivalence of the main result in \cite{ASY2} and
 equation \eqref{eqn:-Vira xi_n} for $n=2$.
 The proofs in \cite{ASY} and \cite{ASY2}
 used Pfaffian expressions for Schur Q-polynomials, which could not be generalized to cases of $n \neq 2$.
 Our proofs of equations \eqref{eqn:Vira xi_n} and \eqref{eqn:-Vira xi_n}  work for all $n \geq 2$, and we will not use Pfaffian expressions for Schur Q-polynomials when $n=2$.

Note that the class of functions $Q_\la^{(n)}$ does not include Schur polynomials $s_\la(\mathbf{t})=Q_\la(\mathbf{t};0)$.
The action of Virasoro operators on $s_\la$ has the following simple form: For all $\la\in\mathbb{Z}^l$ and $m \geq 1$,
\begin{align}\label{eqn:Vira Schur}
L_m^{S} \cdot s_\la =\sum_{i=1}^l \left( \la_i-\frac{2i+m-1}{2} \right) s_{\la-m\epsilon_i},
\end{align}
where
\begin{align}\label{eqn:Vira op Schur}
L_m^{S} := \sum_{k\geq1}kt_k\frac{\partial}{\partial t_{k+m}} +\frac{1}{2}\sum_{k=1}^{m-1} \frac{\partial^2}{\partial t_k\partial t_{m-k}}.
\end{align}
Equation \eqref{eqn:Vira Schur}, or some equivalent forms of it, might have been implicitly used in the study of super-integrability of Gaussian Hermitian matrix models in \cite{MMMR}. However, we could not find a literature where such formulas were explicitly written down. So we will give a proof
of  equation \eqref{eqn:Vira Schur} in Appendix \ref{sec:Schur}, where a formula of $L_m^{S} \cdot  s_\la$ with $m<0$ will also be given.

This paper is organized as follows. In section \ref{sec:pre}, we review the definition of Hall-Littlewood polynomials and prove multiplication formula \eqref{eqn:multi rtr}. In section \ref{sec:xi_n}, we prove Theorems \ref{thm:Vira xi_n} and \ref{thm:-Vira xi_n}. The proof of equation \eqref{eqn:Vira Schur} will be given in Appendix  \ref{sec:Schur}.

\begin{ack}
The authors would like to thank Alexei Morozov for bringing reference \cite{MM21} to our attention, and thank Naihuan Jing, Andrei Mironov, and Alexei Morozov for helpful discussions.
\end{ack}

\section{Preliminaries}\label{sec:pre}

\subsection{Hall-Littlewood polynomials}

In this paper, Hall-Littlewood functions will be considered as polynomials of variables
$\mathbf{t}=(t_1,t_2,...)$, where $r t_r = p_r$ corresponds to the
$r$-th power sum function in the theory of symmetric functions. Hence it is natural to assign $\deg t_r=r$.
The ring $ \Lambda:=\mathbb{C}[\mathbf{t}]$ is isomorphic to the ring of symmetric functions.
Set $\Lambda(\rho):=\Lambda \bigotimes_\mathbb{C} \mathbb{C}(\rho)$ where $\rho$ is a parameter.
A $\mathbb{C}(\rho)$-basis of $\Lambda(\rho)$ is given by $\{ t_\la \mid \la \in \mathcal{P} \}$,
where
$t_\la := \prod_{i=1}^l t_{\la_i}$ for $\la=(\la_1, \cdots, \la_l)$.
Let $l(\la):=l$ if $\la=(\la_1, \cdots, \la_l)$.
There is a natural scalar product on $\Lambda(\rho)$ with values in $\mathbb{C}(\rho)$ given by
\begin{align*}
\langle t_\la, t_\mu\rangle = \delta_{\la,\mu}\frac{z_\la(\rho)}{\left( \prod_{i=1}^{l(\la)} \la_i \right)^2}
\end{align*}
for $\la, \mu \in \mathcal{P}$,
where $\delta_{\la,\mu}$ is the Kronecker symbol,
\[z_\la(\rho) :=\frac{\prod_{k\geq1} k^{m_k(\la)}m_k(\la)!} {\prod_{i=1}^{l(\la)} (1-\rho^{\la_i})} \]
 with $ m_k(\la) :=\# \{ j|\la_j=k \}$.
Using this inner product, we can define the adjoint operator $f^\perp$ of an operator $f$ on $\Lambda(\rho)$ by
\[ \langle f v, \, w \rangle = \langle v, \, f^\perp w \rangle \]
for all $u, v \in  \Lambda(\rho)$.
It is well known that
\begin{align}\label{eqn:adj}
t_r^\perp=\frac{1}{r(1-\rho^r)}\frac{\partial}{\partial t_r},
\end{align}
where $t_r$ is understood as the operator multiplying $t_r$ (see, for example, \cite{JL}).

We will follow the vertex operator realization of Hall-Littlewood polynomials  introduced by Jing in \cite{J}.
Let $B_n$, $n \in \mathbb{Z}$, be the operators on  $\Lambda(\rho)$ whose generating function
$B(u) := \sum_{n\in\mathbb{Z}}B_n u^{n} $ is given by
\begin{align*}
B(u)=\exp\Bigg(\sum_{n\geq1}(1-\rho^n)t_n u^n\Bigg) \exp\Bigg(-\sum_{n\geq1}(1-\rho^n)t_n^\perp u^{-n}\Bigg),
\end{align*}
where $u$ is a parameter. Then $B_n$ is an operator of degree $n$, i.e. $\deg B_n w = \deg w + n$ if
$w$ is a homogeneous element in $\Lambda(\rho)$.
Hall-Littlewood polynomial associated with $\la=(\la_1, \cdots, \la_l) \in\mathbb{Z}^l$ is defined by
\begin{align}\label{eqn:def Q}
Q_\la(\mathbf{t};\rho) :=B_{\la_1}\cdots B_{\la_l} \cdot 1.
\end{align}
It follows that $Q_\la(\mathbf{t};\rho)$ is a homogeneous polynomial of degree $|\la|:=\sum_{i=1}^l \la_i$  and
$Q_\la(\mathbf{t};\rho)=0$
if there exists $1 \leq j \leq l$ such that $\sum_{i=j}^l\la_i<0$. Moreover $Q_{(0)}(\mathbf{t};\rho)=B_0 \cdot 1 = 1$, and
the generating function for
$Q_{(n)}(\mathbf{t};\rho) = B_n \cdot 1$ with $n \geq 0$ is given by
\[ \sum_{n \geq 0} Q_{(n)}(\mathbf{t};\rho) u^n = \exp \left\{ \sum_{n \geq 1} (1-\rho^n) t_n  u^n \right\}. \]

Operators $B_{n}$ satisfy the following relation
\begin{equation}
B_{m-1}B_{n}-\rho B_{m}B_{n-1}=\rho B_{n}B_{m-1}-B_{n-1}B_{m}.
\end{equation}
If $r=n-m$ is a positive odd integer, repeatedly applying the above formula, we have
\begin{align} \label{eqn:BnBmodd}
B_m B_n = \rho B_n B_m + (\rho^2-1) \sum_{i=1}^{\frac{r-1}{2}} \rho^{i-1} B_{n-i} B_{m+i}.
\end{align}
Similarly if $r=n-m$ is a positive even integer, we have
\begin{align} \label{eqn:BnBmeven}
B_m B_n =
\rho B_n B_m +(\rho^2-1)\sum_{i=1}^{\frac{r}{2}-1} \rho^{i-1} B_{n-i} B_{m+i} +
         \rho^{r/2-1}(\rho-1) B_{n-r/2}B_{m+r/2}.
\end{align}

Equations \eqref{eqn:BnBmodd} and \eqref{eqn:BnBmeven} can be used to
write equations for permuting adjacent components of $\la=(\la_1, \cdots, \la_l) \in \mathbb{Z}^l$ in $Q_{\la}$.
For example, if $r=\la_{i+1}-\la_i$ is a positive odd integer, then  by equation \eqref{eqn:BnBmodd}, we have
\begin{align} \label{eqn:relation_of_gQ}
	Q_{(\cdots,\la_i,\la_{i+1},\cdots)}=
	\rho Q_{(\cdots,\la_{i+1},\la_i,\cdots)} +(\rho^2-1)\sum_{k=1}^{(r-1)/2} \rho^{k-1} Q_{(\cdots,\la_{i+1}-k,\la_i+k,\cdots)}
\end{align}
(see, for example, Lemma 1 in \cite{Mor} and Example 2 in \cite{Mac} III.2).
Similar equations can be easily written down using equation \eqref{eqn:BnBmeven} if $r=\la_{i+1}-\la_i$ is a positive even integer.
Together with the fact that $Q_{(\la,0)}=Q_{(\la)}$ and $Q_{\la}=0$ if $\la_{l}<0$,
we can use these equations to represent all $Q_\la$ as linear combinations of
$Q_\mu$ with $\mu \in \mathcal{P}$.

\begin{exa}\label{exa:relation Schur}
	If $\rho=0$, then $Q_{\la}=s_\la$ is the Schur polynomial.
	In this case
	\begin{align}\label{eqn:schur change}
	s_{(...,\la_i,\la_{i+1},...)}=
	\begin{cases}
	-s_{(...,\la_{i+1}-1,\la_i+1,...)},& \text{\ if\ } \la_{i+1} - \la_i \geq 2,\\
	0,& \text{\ if\ } \la_{i+1} - \la_i =1.
	\end{cases}
	\end{align}
\end{exa}

\begin{exa}\label{exa:relation Q}
	If $\rho=-1$, then $Q_\la$ is the Schur Q polynomial. In this case
	\begin{align}\label{eqn:schur Q change}
	Q_{(...,\la_i,\la_{i+1},...)}=
	-Q_{(...,\la_{i+1},\la_i,...)} + \delta_{\la_i, -\la_{i+1}} 2(-1)^{\la_{i}} Q_{(...,\widehat{\la}_i,\widehat{\la}_{i+1},...)},
	\end{align}
     which follows from equations \eqref{eqn:BnBmodd}, \eqref{eqn:BnBmeven},
    and the fact that $B_n^2=\delta_{n,0}$ for $n \in \mathbb{Z}$ when $\rho=-1$.
\end{exa}

 Classical definition of $Q_\la(\mathbf{t};\rho)$ usually requires $\la$ to be a partition (see, for example, \cite{Mac}). However this restriction is not convenient when dealing with derivatives of $Q_\la(\mathbf{t};\rho)$.
 In fact, derivatives of Hall-Littlewood polynomials are given by the following simple formula (see, for example, p144 in \cite{Mor76} and Theorem 2.4 in \cite{JL}):
\begin{align}\label{eqn:derivative}
 \frac{\partial}{\partial t_r} Q_{\la}(\mathbf{t};\rho) = (1-\rho^r) \sum_{i=1}^{l}Q_{\la-r\epsilon_i}(\mathbf{t};\rho)
\end{align}
for all  $\la \in \mathbb{Z}^l$ and $r \geq 1$, where $\la-r\epsilon_i$ is defined by equation \eqref{eqn:la-a}. Note that $\la-r\epsilon_i$ may no be a partition even if $\la$ is. This is one of the reasons we need to extend the definition of $Q_{\la}(\mathbf{t};\rho)$
to allow $\la \in \mathbb{Z}^l$.
Equation \eqref{eqn:derivative} can be obtained by repeatedly applying the following relation (see, for example, equation (2.26) in \cite{JL}):
\begin{align}\label{eqn:t_r^perp B_m}
t_r^\perp B_m = \frac{1}{r}B_{m-r} +B_m t_r^\perp.
\end{align}

In the case that $\rho=\xi_n$ is the $n$-th root of unity, it follows from equation \eqref{eqn:derivative} that $Q_\la(\textbf{t}; \xi_n)$
does not depend on variables $t_{mn}$ for all positive integers $m$.

\subsection{Multiplication formula}
In this subsection, we give a proof for equation \eqref{eqn:multi rtr}  which calculates the product of $p_r:= r t_r$ and Hall-Littlewood polynomials. We need the following
\begin{lem}
For all $r \geq 1$ and $m \in \mathbb{Z}$,
\begin{align}\label{eqn:p_r B_m}
p_r \circ B_m=B_m \circ p_r +B_{m+r},
\end{align}
where "$\circ$" is the composition of operators
\end{lem}
{\bf Proof}:
Note that
\begin{align*}
 B(u)
=&\exp\Bigg(\sum_{n\geq1} \frac{1-\rho^n}{n} p_n u^n \Bigg)  \exp\Bigg(-\sum_{n\geq1} u^{-n} \frac{\partial}{\partial p_n}  \Bigg).
\end{align*}
The operator $p_r$ commutes with all operators on the right hand side of this equation except
$\frac{\partial}{\partial p_r}$.
It is straightforward to check that
\begin{align}
	\exp{\Big(a\frac{\partial}{\partial p_r}\Big)} \circ p_r
	=p_r \circ \exp{\Big(a\frac{\partial}{\partial p_r}\Big)} + a\exp{\Big(a\frac{\partial}{\partial p_r}\Big)}
\end{align}
for all $a$ which does not depend on $p_r$. One can prove this formula by applying both sides of the equation to an arbitrary
function and using Leibniz rule. Hence we have
\begin{align*}
& p_r \circ B(u)  \\
=&\exp\Bigg(\sum_{n\geq1}\frac{1-\rho^n}{n}p_nu^n\Bigg)
            \Bigg\{\exp\Bigg(-\sum_{n\geq1} u^{-n} \frac{\partial}{\partial p_n}  \Bigg) \circ p_r
            +u^{-r}\exp\Bigg(-\sum_{n\geq1} u^{-n} \frac{\partial}{\partial p_n}  \Bigg)\Bigg\}\\
=&B(u) \circ p_r +u^{-r} B(u).
\end{align*}
Since $B(u)=\sum_{m \in \mathbb{Z}} B_m u^m$,  the coefficient of $u^m$ in the above equation gives the desired  formula.
$\Box$

Now we give a more precise statement of equation \eqref{eqn:multi rtr}:
\begin{pro} \label{pro:multipr}
For all $r \geq 1$ and $\la\in\mathbb{Z}^l$, we have
\begin{align*}
p_r Q_{\la}=\sum_{i=1}^{l}Q_{\la+r\epsilon_i}
+\sum_{\mu \in \mathcal{P} \atop|\mu|=r}c_\mu(\rho)Q_{(\la,\mu)},
\end{align*}
where
\begin{align}\label{eqn:c}
c_\mu(\rho) = \frac{\rho^{n(\mu)}\phi_{l(\mu)-1}(\rho^{-1})}{b_\mu(\rho)}
\end{align}
with
\begin{align*}
n(\mu):=\sum_{i=1}^{l(\mu)}(i-1)\mu_i,\  \phi_k(\rho):=\prod_{i=1}^k (1-\rho^i),\\
m_i(\mu):=\#\{j|\mu_j=i\}, \ b_\mu(\rho):=\prod_{i=1}^\infty\phi_{m_i(\mu)}(\rho).
\end{align*}
\end{pro}
{\bf Proof}:
For $\la=(\la_1, \cdots, \la_l) \in \mathbb{Z}^l$, repeatedly applying equation \eqref{eqn:p_r B_m}, we have
\begin{align}\label{eqn:proof pr}
p_r \circ B_{\la_1} \cdots B_{\la_l}
=&\sum_{i=1}^{l}B_{\la_1}\cdots B_{\la_i+r}\cdots B_{\la_l} + B_{\la_1}\cdots B_{\la_l}\circ p_r .
\end{align}
The proposition is then obtained by applying both sides of this equation to $1$ and using the formula
\begin{align} \label{eqn:pr}
p_r = \sum_{\mu \in \mathcal{P} \atop|\mu|=r} c_\mu(\rho) Q_{\mu}.
\end{align}
Note that equation \eqref{eqn:pr} follows from equation (7.1) and Example 2 in Section III.7 in \cite{Mac}.
$\Box$

\begin{exa} \label{exa:c Schur}
	For the Schur polynomial case, $\rho=0$. The coefficient
    $c_\la(0)$ should be understood as
	\begin{align*}
	c_\la(0)=\lim_{\rho \rightarrow 0} \frac{\rho^{n(\la)}\phi_{l(\la)-1}(\rho^{-1})}{b_\la(\rho)}=
	\begin{cases}
	(-1)^{l(\la)-1},& \text{\ if\ }  \la=(k,1,...,1) \text{\ with \ }  k\geq1,\\
	0,& \text{\ otherwise\ }.
	\end{cases}
	\end{align*}
	Together with equation \eqref{eqn:schur change}, one can use Proposition \ref{pro:multipr} to recover the following well-known formula
	\begin{align}\label{eqn:MN Schur}
	p_r s_\la=\sum_{\mu\setminus\la\text{\ is\ a\ border\ strip\ of\ size\ }r}(-1)^{ht(\mu\setminus\la)} s_\mu
	\end{align}
     for $\la \in \mathcal{P}$ (c.f. Example 11 in Section I.3 in \cite{Mac}). As pointed out in Example 5 in Section I.7 in \cite{Mac},
     this formula implies the well-known Murnaghan-Nakayama rule for calculating characters of irreducible representations of the symmetric group.
\end{exa}

\begin{exa}\label{exa:c n=2}
	For the Schur Q-polynomial case, $\rho=-1$.  In this case $Q_\la$ does not depend on $p_{2k}$ for all $k \geq 1$
and $Q_{\la}=0$ if $\la \in \mathcal{P}$ is not strict.  Moreover $\phi_k(-1)=0$ if $k \geq 2$, and
	\begin{align*}
	c_\la(-1)=
	\begin{cases}
	\frac{(-1)^{m}}{2},& \text{\ if\ } \la=(k,m) \text{\ with \ }  k \geq m \geq 0,\\
	0,& \text{\ if \ } l(\la) \geq 3.
	\end{cases}
	\end{align*}
	Thus formula \eqref{eqn:multi rtr} in this case agrees with previous results in \cite{B} and \cite{LY}, which were
 proved using different methods.
\end{exa}

\begin{exa}
	For $\rho=\xi_n$ with $n \geq 2$, $\phi_k(\rho)=\phi_k(\rho^{-1})=0$ if $k \geq n$. Hence $c_{\la}(\xi_n)=0$ if $l(\la) \geq n+1$.
\end{exa}

\section{Action of Virasoro operators on Hall-Littlewood polynomials}
\label{sec:xi_n}

In this section, we prove Theorems \ref{thm:Vira xi_n} and \ref{thm:-Vira xi_n} by induction on the length of $\la$.
Fix an integer $n \geq 2$.
To carry out the induction process, we need to compute the commutator $[L_{m}^{(n)}, B_k]$, where $L_{m}^{(n)}$ is
defined by equation \eqref{eqn:Vira op xi_n}.
We first compute $[\widehat{L}_{m}^{(n)}, B_k]$ where
\begin{equation} \label{eqn:Lhat}
\widehat{L}_{m}^{(n)} := \sum_{k \geq 1} k t_k \frac{\partial}{\partial t_{k+mn}}.
\end{equation}
Recall $p_k=k t_k$ for all $k$.  Operators $\widehat{L}_{m}^{(n)}$ can also be written as
\begin{equation} \label{eqn:Lhat2}
\widehat{L}_{m}^{(n)} := \sum_{k \geq \iota_m} (1-\rho^{k+mn}) p_k p_{k+mn}^\perp,
\end{equation}
where
\begin{align}
\iota_m :=
\begin{cases}
	1,& \text{\ if\ }    m \geq 0,\\
	1-mn,& \text{\ if\ }    m < 0.
	\end{cases}
\end{align}
For any integer $N > \iota_m$, define
\begin{equation}
[\widehat{L}_{m}^{(n)}]_N := \sum_{k = \iota_m}^N (1-\rho^{k+mn}) p_k p_{k+mn}^\perp.
\end{equation}
Recall that we have used the symbol  "$\circ$" for composition of operators. This symbol may be dropped
if there is no ambiguity about the product of two operators.

We will need the following two lemmas to calculate $[\widehat{L}_{m}^{(n)}, B_r]$
\begin{lem} \label{lem:LNBr}
For any $r \in \mathbb{Z}$,
\begin{equation} \label{eqn:LNBr}
 [\widehat{L}_{m}^{(n)}]_N \circ B_r = B_r \circ [\widehat{L}_{m}^{(n)}]_N
            + \sum_{k = \iota_m}^N (1-\rho^{k+mn}) \left\{ B_{r-mn} + B_{r+k} \circ p_{k+mn}^\perp + B_{r-k-mn} \circ p_k \right\}.
 \end{equation}
\end{lem}
{\bf Proof}:
Using equations \eqref{eqn:t_r^perp B_m} and \eqref{eqn:p_r B_m}, we obtain
\begin{equation}
p_k p_j^\perp \circ B_r
= B_r \circ p_k p_j^\perp + B_{r-j+k} + B_{r+k} \circ p_j^\perp + B_{r-j} \circ p_k  ,
\end{equation}
for all $r \in \mathbb{Z}$ and $j, k \geq 1$. This implies the lemma.
$\Box$

\begin{lem} \label{lem:iden}
If $N > \max\{a, a+r\}$, then for any homogeneous polynomial $f(\textbf{t})$ of degree $a$, we have
\begin{align}\label{eqn:iden}
	 \sum_{k=1}^{N} (1-\rho^k) B_{r-k}\circ p_k \cdot f  
	= \left\{ r- \sum_{k=1}^N (1-\rho^k) \right\} B_r \cdot f
	 -\sum_{k=1}^{N} (1-\rho^k) B_{r+k} \circ p_k^\perp \cdot f
	\end{align}
for all $r \in \mathbb{Z}$ and $a \geq 0$.
\end{lem}
{\bf Proof}:
Note that
\[ \widehat{L}_{0}^{(n)} = \sum_{k \geq 1} k t_k \frac{\partial}{\partial t_{k}} \hspace{10pt}  \text{\ and \ } \hspace{10pt}
[\widehat{L}_{0}^{(n)}]_N = \sum_{k=1}^N k t_k \frac{\partial}{\partial t_{k}} .\]
For any homogeneous polynomial $g(\textbf{t})$ of degree less than $N$, $\frac{\partial g}{\partial t_{k}} =0$ for all $k>N$. Hence
\[  [\widehat{L}_{0}^{(n)}]_N \cdot g = \widehat{L}_{0}^{(n)} \cdot g = \deg (g) \cdot  g.\]
In particular, since $f$ and $B_r \cdot f$ are homogeneous of degree $a$ and $a+r$ respectively,
we have
\[ [\widehat{L}_{0}^{(n)}]_N \cdot f = a f \hspace{10pt}  \text{\ and \ } \hspace{10pt}
[\widehat{L}_{0}^{(n)}]_N \circ B_r \cdot f = (a+r) B_r \cdot f. \]
Applying both sides of equation \eqref{eqn:LNBr} with $m=0$ to function $f$, we obtain the desired formula.
$\Box$

Now we are ready to compute $[\widehat{L}_{m}^{(n)}, B_r]$.
\begin{pro} \label{pro:LhB}
Assume $\rho=\xi_n$. Then for all $r \in \mathbb{Z}$,
\begin{equation} \label{eqn:LBm>0}
[\widehat{L}_{m}^{(n)}, B_r] = (r-mn) B_{r-mn} - \sum_{k=1}^{mn} (1-\xi_n^k) B_{r-mn+k} \circ p_k^\perp
\end{equation}
if $ m \geq 1 $, and
\begin{equation} \label{eqn:LBm<0}
[\widehat{L}_{m}^{(n)}, B_r] = r B_{r-mn} - \sum_{k=1}^{-mn} (1-\xi_n^k) B_{r-mn-k} \circ p_k
\end{equation}
if $ m \leq -1$.
\end{pro}
{\bf Proof}:
We only need to show that equalities hold when both sides of equations \eqref{eqn:LBm>0} and \eqref{eqn:LBm<0}
acting on an arbitrary homogeneous polynomial $f(\textbf{t})$ of degree $a \geq 0$.
Choose an integer $N > \max\{a, a+r\}+\iota_m$. Then
\[ [\widehat{L}_{m}^{(n)}, B_r] \cdot f = ([\widehat{L}_{m}^{(n)}]_N \circ B_r - B_r \circ [\widehat{L}_{m}^{(n)}]_N ) \cdot f. \]
So by Lemma~\ref{lem:LNBr}, we have
\begin{eqnarray} \label{eqn:LmBrf}
[\widehat{L}_{m}^{(n)}, B_r] \cdot f
&=&  \sum_{k = \iota_m}^N (1-\xi_n^{k}) \left\{ B_{r-mn} + B_{r+k} \circ p_{k+mn}^\perp + B_{r-k-mn} \circ p_k \right\} \cdot f
\end{eqnarray}
Here we have used the fact $\xi_n^{k \pm mn}=\xi_n^{k}$ since $\xi_n$ is the $n$-th root of unity.
Applying Lemma \ref{lem:iden} to the last term on the right hand side of equation \eqref{eqn:LmBrf}, we have
\begin{eqnarray} \label{eqn:appIden}
&& \sum_{k = \iota_m}^N (1-\xi_n^{k})  B_{r-k-mn} \circ p_k  \cdot f
+ \sum_{k =1}^{\iota_m-1} (1-\xi_n^{k})  B_{r-k-mn} \circ p_k  \cdot f  \nonumber \\
&=&  \left\{ r-mn - \sum_{k=1}^N (1-\xi_n^k) \right\} B_{r-mn} \cdot f
	 -\sum_{k=1}^{N} (1-\xi_n^k) B_{r-mn+k} \circ p_k^\perp \cdot f.
\end{eqnarray}

If $m \geq 1$, then $\iota_m=1$. Combination of equations \eqref{eqn:LmBrf} and \eqref{eqn:appIden} gives
the following
\begin{eqnarray}
[\widehat{L}_{m}^{(n)}, B_r] \cdot f
&=&  \left( r-mn  \right) B_{r-mn} \cdot f
    -\sum_{k=1}^{mn} (1-\xi_n^k) B_{r-mn+k} \circ p_k^\perp \cdot f \nonumber \\
 &&   + \sum_{k=N+1}^{N+mn} (1-\xi_n^k) B_{r-mn+k} \circ p_k^\perp \cdot f. \nonumber
\end{eqnarray}
The last term on the right hand side of this equation vanishes since $p_k^\perp \cdot f = 0$
for $k>N> \deg (f)$. This proves equation \eqref{eqn:LBm>0} since $f$ is an arbitrary homogenous
polynomial.

If $m \leq -1$, then $\iota_m=1-mn$. Combination of equations \eqref{eqn:LmBrf} and \eqref{eqn:appIden} gives
the following
\begin{eqnarray}
[\widehat{L}_{m}^{(n)}, B_r] \cdot f
&=&  \left\{ r-mn - \sum_{k=1}^{-mn} (1-\xi_n^k) \right\} B_{r-mn} \cdot f
    -\sum_{k =1}^{-mn} (1-\xi_n^{k})  B_{r-k-mn} \circ p_k  \cdot f  \nonumber \\
 &&   - \sum_{k=N+mn+1}^{N} (1-\xi_n^k) B_{r-mn+k} \circ p_k^\perp \cdot f. \nonumber
\end{eqnarray}
The last term on the right hand side of this equation vanishes since $p_k^\perp \cdot f = 0$
for $k>N+mn> \deg (f)$. Moreover
\[ \sum_{k=1}^{-mn} (1-\xi_n^k) = -mn - \frac{\xi_n-\xi_n^{-mn+1}}{1-\xi_n}=-mn\]
since $\xi_n$ is the $n$-th root of unity.
This proves equation \eqref{eqn:LBm<0} since $f$ is an arbitrary homogenous
polynomial.
The proposition is thus proved.
$\Box$

\begin{cor}
For $\rho=\xi_n$, $m \geq 1$, and $r \in \mathbb{Z}$,
    \begin{equation} \label{eqn:LtBm>0}
    [\widetilde{L}_{m}^{(n)}, B_r] = r B_{r-mn} + \sum_{k=1}^{mn} (1-\xi_n^{-k}) B_{r-mn+k} \circ p_k^\perp
    \end{equation}
where
    \begin{align} \label{eqn:Ltilde}
	\widetilde{L}_{m}^{(n)}
    &:= \sum_{k \geq 1} k t_k \frac{\partial}{\partial t_{k+nm}}
	+\frac{1}{2}\sum_{k=1}^{mn-1} \frac{\partial^2}{\partial t_k \partial t_{mn-k}}.
	\end{align}
\end{cor}
{\bf Proof}:
For $m \geq 0$, let
\begin{align*} 
	W_{m}^{(n)}
    &:= \sum_{k=1}^{mn-1} \frac{\partial^2}{\partial t_k \partial t_{mn-k}}
    = \sum_{k=1}^{mn-1} (1-\rho^k)(1-\rho^{mn-k}) p_k^\perp p_{mn-k}^\perp.
	\end{align*}
Then \[\widetilde{L}_{m}^{(n)}=\widehat{L}_{m}^{(n)} + \frac{1}{2} W_{m}^{(n)}.\]
Using equation \eqref{eqn:t_r^perp B_m} twice, we obtain
\begin{equation}
[p_k^\perp p_{j}^\perp,  B_{r}]
=  B_{r-j-k} +B_{r-j} p_k^\perp + B_{r-k} p_{j}^\perp
\end{equation}
for all $r \in \mathbb{Z}$ and $j,k \geq 1$. Hence
\begin{equation} \label{eqn:WBm>0rho}
[W_{m}^{(n)},  B_{r}]
=  \sum_{k=1}^{mn-1} (1-\rho^k)(1-\rho^{mn-k}) \left\{ B_{r-mn} + 2 B_{r-mn+k} \, p_k^\perp  \right\}.
\end{equation}
If $\rho = \xi_n$, then $\rho^{mn-k}=\xi_n^{-k}$, $\sum_{k=1}^{mn} \xi_n^k =  \sum_{k=1}^{mn} \xi_n^{-k} = 0$, and
 \[ \sum_{k=1}^{mn-1} (1-\rho^k)(1-\rho^{mn-k}) = \sum_{k=1}^{mn} (2-\xi_n^k - \xi_n^{-k}) = 2mn. \]
 Hence
 \begin{equation}
[W_{m}^{(n)},  B_{r}]
=  2mn B_{r-mn} +  2 \sum_{k=1}^{mn} (1-\xi_n^k)(1-\xi_n^{-k})   B_{r-mn+k} \, p_k^\perp  .
\end{equation}
The corollary then follows from this equation and equation \eqref{eqn:LBm>0}.
$\Box$

We are now ready to prove the first main result of this paper.

{\bf Proof of Theorem~\ref{thm:Vira xi_n}}:
If $\rho=\xi_n$, $Q_\la(\textbf{t}; \rho) = Q_{\la}^{(n)}$ does not depend on variables $t_{mn}$ for all $m \geq 1$.
Hence \[ L_{m}^{(n)} Q_{\la}^{(n)}=\widetilde{L}_{m}^{(n)} Q_{\la}^{(n)}, \]
where $\widetilde{L}_{m}^{(n)}$ is defined by equation \eqref{eqn:Ltilde}.
For $m=0$, equation \eqref{eqn:Vira xi_n} follows from the fact that $ Q_{\la}^{(n)}$ is homogeneous of degree $|\la|$.

For $m>0$,  we prove this theorem by induction on $l(\la)$.
If $l(\la)=1$ and $\la=(\la_1)$, then
\[
L_{m}^{(n)} Q_{\la}^{(n)} = \widetilde{L}_{m}^{(n)} B_{\la_1} \cdot 1 = \la_1 B_{\la_1-mn} \cdot 1 = \la_1 Q_{\la - mn \epsilon_1}^{(n)},
\]
where the second equality follows from equation \eqref{eqn:LtBm>0} since $\widetilde{L}_{m}^{(n)} \cdot 1 = p_k^\perp \cdot 1 =0$.
Hence equation \eqref{eqn:Vira xi_n} is true when $l(\la)=1$.

If $l(\la)=l > 1$, let $\la=(\la_1, \widehat{\la})$ where $\widehat{\la} = (\la_2, \cdots, \la_l) $.
By equations \eqref{eqn:adj} and \eqref{eqn:derivative},
\begin{equation} \label{eqn:derivp}
p_k^\perp \cdot Q_\la = \sum_{i=1}^{l} Q_{\la-k \epsilon_i}
\end{equation}
for all $\la \in \mathbb{Z}^l$ and $k \geq 1$.
By equation \eqref{eqn:LtBm>0},
\begin{eqnarray*}
L_{m}^{(n)} Q_{\la}^{(n)} &=& \widetilde{L}_{m}^{(n)} B_{\la_1} \cdot Q_{\widehat{\la}}^{(n)} 
= \left\{ B_{\la_1} \widetilde{L}_{m}^{(n)} + \la_1 B_{\la_1 -mn} + \sum_{k=1}^{mn} (1-\xi_n^{-k}) B_{\la_1-mn+k} \circ p_k^\perp \right\} \cdot Q_{\widehat{\la}}^{(n)} \nonumber \\
&=& B_{\la_1} L_{m}^{(n)} Q_{\widehat{\la}}^{(n)} + \la_1 Q_{\la-mn \epsilon_1}^{(n)}
    + \sum_{k=1}^{mn-1} (1-\xi_n^{-k}) \sum_{i=2}^l Q_{\la-(mn-k) \epsilon_1- k \epsilon_i}^{(n)}.
\end{eqnarray*}
By induction hypothesis,
\begin{eqnarray*}
B_{\la_1} L_{m}^{(n)} Q_{\widehat{\la}}^{(n)}
&=& \sum_{i=2}^l \la_i Q_{\la - mn \epsilon_i}^{(n)}
    + \sum_{k=1}^{mn-1} \sum_{i,j=2 \atop i>j}^l (1-\xi_n^{-k}) Q_{\la - k \epsilon_i - (mn-k) \epsilon_j}^{(n)}.
\end{eqnarray*}
Combination of the above two equations gives equation \eqref{eqn:Vira xi_n}.
This finishes the proof of Theorem~\ref{thm:Vira xi_n}.
$\Box$

Before proving Theorem \ref{thm:-Vira xi_n}, we first prove the following simpler formula
\begin{thm}\label{thm:-Vira' xi_n}
For any $n\geq2$, $m \geq 1$, $\la\in\mathbb{Z}^l$, we have
\begin{align}\label{eqn:-Vira' xi_n}
\widehat{L}_{-m}^{(n)} \cdot Q^{(n)}_\la
=&\sum_{i=1}^l \la_i Q^{(n)}_{\la+mn\epsilon_i}
 + \sum_{k=1}^{mn}\sum_{i,j=1\atop i>j}^{l} (\xi_n^k-1)Q^{(n)}_{\la+k\epsilon_i+(mn-k)\epsilon_j} \nonumber\\
 & + \sum_{i=1}^l \sum_{k=1}^{mn} (\xi_n^{k}-1) \sum_{\mu \in \mathcal{P} \atop |\mu|=k} c_\mu(\xi_n) Q^{(n)}_{(\la+(mn-k)\epsilon_i,\mu)},
\end{align}
where $\widehat{L}_{-m}^{(n)}$ is defined by equation \eqref{eqn:Lhat}.
\end{thm}
{\bf Proof:}
If $l(\la)=l \geq 1$, let $\la=(\la_1, \widehat{\la}) \in \mathbb{Z}^l$ where $\widehat{\la} = (\la_2, \cdots, \la_l) $.
Then we have
\begin{eqnarray}
\widehat{L}_{-m}^{(n)} \cdot Q^{(n)}_\la
&=& \widehat{L}_{-m}^{(n)} B_{\la_1} \cdot  Q^{(n)}_{\widehat{\la}} \nonumber \\
&=& \left\{ B_{\la_1} \widehat{L}_{-m}^{(n)}+\la_1 B_{\la_1+mn}
                + \sum_{k=1}^{mn} (\xi_n^k-1) B_{\la_1-k+mn} \circ p_k \right\} \cdot Q^{(n)}_{\widehat{\la}} \nonumber \\
&=& B_{\la_1} \widehat{L}_{-m}^{(n)}  Q^{(n)}_{\widehat{\la}} + \la_1 Q^{(n)}_{\la+mn\epsilon_1}  \nonumber \\
&&    + \sum_{k=1}^{mn} (\xi_n^k-1) B_{\la_1-k+mn}
        \left\{ \sum_{i=1}^{l-1}  Q^{(n)}_{\widehat{\la}+ k \epsilon_i}
        + \sum_{\mu \in \mathcal{P} \atop |\mu|=k} c_\mu (\xi_n) Q_{(\widehat{\la},\mu)}^{(n)} \right\} \nonumber \\
&=& B_{\la_1} \widehat{L}_{-m}^{(n)}  Q^{(n)}_{\widehat{\la}} + \la_1 Q^{(n)}_{\la+mn\epsilon_1}  \nonumber \\
 &&   + \sum_{k=1}^{mn} (\xi_n^k-1)
        \left\{ \sum_{i=2}^{l}  Q^{(n)}_{\la + (mn-k) \epsilon_1 + k \epsilon_i}
        + \sum_{\mu \in \mathcal{P} \atop |\mu|=k} c_\mu (\xi_n) Q_{(\la + (mn-k) \epsilon_1,\mu)}^{(n)} \right\} ,
        \label{eqn:vir-ind}
\end{eqnarray}
where the second equality follows from equation \eqref{eqn:LBm<0}, the third equality follows from
equation \eqref{eqn:multi rtr}.

We prove the theorem by induction on $l(\la)$. If $l(\la)=1$, then $\widehat{\la}$ is the empty partition.
In this case
$Q^{(n)}_{\widehat{\la}}=1$ and $\widehat{L}_{-m}^{(n)}  Q^{(n)}_{\widehat{\la}}=0$.
Equation \eqref{eqn:vir-ind} coincides with equation \eqref{eqn:-Vira' xi_n}.
Hence the theorem holds for $l(\la)=1$.

If $l(\la)>1$, by induction hypothesis,
\begin{eqnarray}
B_{\la_1} \widehat{L}_{-m}^{(n)}  Q^{(n)}_{\widehat{\la}}
&=& \sum_{i=2}^l \la_i Q^{(n)}_{\la+mn\epsilon_i}
 + \sum_{k=1}^{mn}\sum_{i,j=2 \atop i>j}^{l} (\xi_n^k-1)Q^{(n)}_{\la+k\epsilon_i+(mn-k)\epsilon_j} \nonumber\\
&& + \sum_{i=2}^l \sum_{k=1}^{mn} (\xi_n^{k}-1) \sum_{\mu \in \mathcal{P} \atop |\mu|=k} c_\mu(\xi_n)
    Q^{(n)}_{(\la+(mn-k)\epsilon_i,\mu)}. \label{eqn:v-indhyp}
\end{eqnarray}
Combining equations \eqref{eqn:vir-ind} and \eqref{eqn:v-indhyp}, we obtain equation \eqref{eqn:-Vira' xi_n}.
Hence the theorem is proved.
$\Box$

Now we are ready to prove the second main theorem of this paper.

{\bf Proof of Theorem \ref{thm:-Vira xi_n}:}
Let
\[ V_m^{(n)} := \sum_{k=1 \atop n \nmid k}^{mn-1} p_k p_{mn-k} \]
for $m \geq 1$.
Since $Q^{(n)}_\la$ does not depend on $t_{kn}$ for all $k \geq 1$, we have
\[ L_{-m}^{(n)} \cdot Q^{(n)}_\la = \widehat{L}_{-m}^{(n)} \cdot Q^{(n)}_\la + \frac{1}{2} V_m^{(n)} Q^{(n)}_\la
\]
for all $\la$. Applying equation \eqref{eqn:multi rtr} twice, we have
\begin{eqnarray}
V_m^{(n)} Q^{(n)}_\la
&=& \sum_{k=1, \atop n\nmid k}^{mn-1}\Big(\sum_{\mu \in \mathcal{P} \atop |\mu|=k}\sum_{j=1}^{l(\mu)} c_{\mu}(\xi_n) Q^{(n)}_{(\la,\mu+(mn-k)\epsilon_j)}
+\sum_{\mu,\nu \in \mathcal{P} \atop |\mu|=k,|\nu|=mn-k} c_{\mu}(\xi_n) c_{\nu}(\xi_n) Q^{(n)}_{(\la,\mu,\nu)}\Big) \nonumber \\
&& + m(n-1)\sum_{i=1}^{l(\la)} Q^{(n)}_{\la+mn\epsilon_i}
+2\sum_{k=1,\atop n\nmid k}^{mn-1}\sum_{i,j=1\atop i>j}^{l(\la)} Q^{(n)}_{\la+k\epsilon_i+(mn-k)\epsilon_j} \nonumber \\
&& +2\sum_{k=1, \atop n\nmid k}^{mn-1} \sum_{i=1}^{l(\la)}  \sum_{\mu \in \mathcal{P} \atop |\mu|=k}
        c_{\mu}(\xi_n)Q^{(n)}_{(\la+(mn-k)\epsilon_i,\mu)}.  \label{eqn:VmQ}
\end{eqnarray}
The theorem then follows from equations \eqref{eqn:-Vira' xi_n} and \eqref{eqn:VmQ}.
$\Box$

\appendix
\vspace{30pt}
\hspace{160pt} {\bf \Large Appendix}

\section{Action of Virasoro operators on Schur polynomials}
\label{sec:Schur}

In this appendix, we study the action of Virasoro operators on Schur polynomials.
These Virasoro operators are given by
\begin{align} \label{eqn:VirS}
L_m^{S}:=\sum_{k\geq1}kt_k\frac{\partial}{\partial t_{k+m}}
+\frac{1}{2}\sum_{k=1}^{m-1}\frac{\partial^2}{\partial t_k\partial t_{m-k}}
+\frac{1}{2}\sum_{k=1}^{-m-1} k(-m-k)t_k t_{-m-k}.
\end{align} \label{eqn:VirShat}
The first order derivative part of this operator is
\begin{equation}
\widehat{L}_m^S = \sum_{k\geq1}kt_k\frac{\partial}{\partial t_{k+m}}.
\end{equation}
$\widehat{L}_m^S$ is just the operator $\widehat{L}_m^{(n)}$ defined in equation \eqref{eqn:Lhat} with $n=1$.
Hence the same proof for Proposition \ref{pro:LhB} with $\rho=0$ and $n=1$ shows the following
\begin{lem}
Assume $\rho=0$. Then for all $r \in \mathbb{Z}$,
\begin{equation} \label{eqn:LShBm>0}
[\widehat{L}_{m}^{S}, B_r] = (r-m) B_{r-m} - \sum_{k=1}^{m}  B_{r-m+k} \circ p_k^\perp
\end{equation}
if $ m \geq 1 $, and
\begin{equation} \label{eqn:LShBm<0}
[\widehat{L}_{m}^{S}, B_r] = r B_{r-m} - \sum_{k=1}^{-m}  B_{r-m-k} \circ p_k
\end{equation}
if $ m \leq -1$.
\end{lem}
Consequently, we have
\begin{cor}
Assume $\rho=0$. Then for all $r \in \mathbb{Z}$,
\begin{equation} \label{eqn:LSBm>0}
[L_{m}^{S}, B_r] = \left( r-\frac{m+1}{2} \right) B_{r-m} -  B_{r} \circ p_m^\perp
\end{equation}
if $ m \geq 1 $, and
\begin{equation} \label{eqn:LSBm<0}
[L_{m}^{S}, B_r] = \left( r-\frac{m+1}{2} \right) B_{r-m} -   B_{r} \circ p_{-m}
\end{equation}
if $ m \leq -1$.
\end{cor}
{\bf Proof}:
Define
\begin{equation}
W_m^S =\sum_{k=1}^{m-1}\frac{\partial^2}{\partial t_k\partial t_{m-k}}
\end{equation}
for $m \geq 0$, and
\begin{equation}
W_m^S= \sum_{k=1}^{-m-1} p_k p_{-m-k}
\end{equation}
for $m<0$.
Then
\[ L_{m}^{S}=\widehat{L}_{m}^{S} + \frac{1}{2} W_m^S.\]

By equation \eqref{eqn:WBm>0rho} with $\rho=0$ and $n=1$, we have
\begin{equation}
[W_m^S, B_r]=(m-1) B_{r-m} + 2 \sum_{k=1}^{m-1} B_{r-m+k} \circ p_k^\perp
\end{equation}
for $m>0$. This equation and equation \eqref{eqn:LShBm>0} imply equation \eqref{eqn:LSBm>0}.

Applying equation \eqref{eqn:p_r B_m} twice, we have
\begin{equation}
p_k p_j \circ B_r = B_r \circ p_k p_j + B_{r+k} \circ p_j + B_{m+j} \circ p_k  + B_{r+j+k}
\end{equation}
for all $r \in \mathbb{Z}$ and $j, k \geq 1$.
Consequently we have
\begin{equation}
[W_m^S, B_r] = (-m-1) B_{r-m}  + 2 \sum_{k=1}^{-m-1}  B_{r-m-k} \circ p_k
\end{equation}
for $m<0$. This equation and equation \eqref{eqn:LShBm<0} imply equation \eqref{eqn:LSBm<0}.
$\Box$


Now we are ready to prove equation \eqref{eqn:Vira Schur}, which we restate as the following theorem:
\begin{thm} \label{thm:VirSchurm>0}
For $m \geq 1$ and $\la \in \mathbb{Z}^l$,
\begin{align*}
L_m^{S} \cdot s_\la =\sum_{i=1}^l \left( \la_i-\frac{2i+m-1}{2} \right) s_{\la-m\epsilon_i}.
\end{align*}
\end{thm}
{\bf Proof}:
We will prove this formula by induction on $l$.
Write $\la = (\la_1, \widehat{\la})$ where $\widehat{\la}=(\la_2, \cdots \la_l)$. Then
by equation \eqref{eqn:LSBm>0}, we have
\begin{eqnarray*}
L_m^S \cdot s_\la &=& L_m^S B_{\la_1} \cdot s_{\widehat{\la}}
    = \left\{ B_{\la_1} L_m^S + \left( \la_1 - \frac{m+1}{2} \right) B_{\la_1 - m} - B_{\la_1} \circ p_m^\perp \right\} \cdot
        s_{\widehat{\la}}.
\end{eqnarray*}

If $l=1$, $\widehat{\la}$ is the empty partition and $s_{\widehat{\la}}=1$.
So $L_m^S \cdot s_{\widehat{\la}} = p_m^\perp \cdot s_{\widehat{\la}} = 0$ and
$L_m^S \cdot s_\la = \left( \la_1 - \frac{m+1}{2} \right) s_{\la-m\epsilon_1}$. Hence the theorem holds for $l=1$.

If $l > 1$, by induction hypothesis and equation \eqref{eqn:derivp}, we have
\begin{eqnarray*}
L_m^S \cdot s_\la
&=& \sum_{i=2}^l \left\{ \la_i - \frac{2(i-1)+m-1}{2} \right\} s_{\la-m\epsilon_i}
    + \left( \la_1 - \frac{m+1}{2} \right) s_{\la-m\epsilon_1}
    -  \ \sum_{i=2}^l  s_{\la-m\epsilon_i}  \\
&=& \sum_{i=1}^l \left( \la_i-\frac{2i+m-1}{2} \right) s_{\la-m\epsilon_i}.
\end{eqnarray*}
The theorem is thus proved.
$\Box$

For the action of the negative branch of Virasoro operators on Schur polynomials, we have the following
\begin{thm}
For  $m \geq 1$ and $\la \in \mathbb{Z}^l$,
\begin{align}\label{eqn:-Vira Schur}
L_{-m}^{S}\cdot s_{\la}
=&\sum_{i=1}^l \left( \la_i-i+\frac{m+1}{2} \right) s_{\la+m\epsilon_i}
-\sum_{k=1}^m (-1)^{m-k} \left( l-k+\frac{m+1}{2} \right) s_{(\la,k,1^{m-k})},
\end{align}
where $1^{j}:=(1, \cdots , 1) \in \mathbb{Z}^{j}$ for $j \geq 0$.
\end{thm}
{\bf Proof}:
We first note that if $l=0$, then $\la$ is an empty partition, $s_\la=1$ and equation \eqref{eqn:-Vira Schur} becomes
\begin{equation} \label{eqn:VirS1}
L_{-m}^{S}\cdot 1
= \sum_{k=1}^m (-1)^{m-k+1} \left(-k+\frac{m+1}{2} \right) s_{(k,1^{m-k})}.
\end{equation}
The proof of this theorem is divided into proofs of statements in the following three steps.

\vspace{10pt}
{\bf Step 1}: For each $m \geq 1$, equation \eqref{eqn:-Vira Schur} follows from equation \eqref{eqn:VirS1}.
\vspace{10pt}

We prove this statement by induction on $l$. Equation \eqref{eqn:VirS1} is the base case with $l=0$ for the induction.
For $l \geq 1$, we write
$\la = (\la_1, \widehat{\la})$ where $\widehat{\la}=(\la_2, \cdots \la_l)$. Then
by equation \eqref{eqn:LSBm<0}, we have
\begin{eqnarray*}
L_{-m}^S \cdot s_\la &=& L_{-m}^S B_{\la_1} \cdot s_{\widehat{\la}}
    = \left\{ B_{\la_1} L_{-m}^S + \left( \la_1 - \frac{-m+1}{2} \right) B_{\la_1 + m} - B_{\la_1} \circ p_{m} \right\} \cdot
        s_{\widehat{\la}}.
\end{eqnarray*}
By Example \ref{exa:c Schur},  equation \eqref{eqn:multi rtr} becomes
\begin{equation} \label{eqn:multiSchur}
p_r s_\mu = \sum_{i=1}^{l(\mu)} s_{\mu + r \epsilon_i} + \sum_{k=1}^r (-1)^{r-k} s_{(\mu, k, 1^{r-k})}
\end{equation}
for all $r \geq 1$ and $\mu \in \mathbb{Z}^{l(\mu)}$.
Using this equation and the induction hypothesis, we obtain
\begin{eqnarray*}
&& L_{-m}^S \cdot s_\la \\
&=& \sum_{i=2}^l \left\{ \la_i - (i-1) + \frac{m+1}{2} \right\} s_{\la+m \epsilon_i}
            - \sum_{k=1}^m (-1)^{m-k} \left(l-1-k+\frac{m+1}{2} \right) s_{(\la, k, 1^{m-k})} \\
&&  +  \left( \la_1 - \frac{-m+1}{2} \right) s_{\la + m \epsilon_1}
    - \left\{ \sum_{i=2}^l s_{\la+m\epsilon_i} + \sum_{k=1}^m (-1)^{m-k} s_{(\la, k, 1^{m-k})} \right\} \\
&=& \sum_{i=1}^l \left\{ \la_i - i + \frac{m+1}{2} \right\} s_{\la+m \epsilon_i}
            - \sum_{k=1}^m (-1)^{m-k} \left(l-k+\frac{m+1}{2} \right) s_{(\la, k, 1^{m-k})} .
\end{eqnarray*}
This completes the proof for the statement in step 1.

\vspace{10pt}
{\bf Step 2}: Equation \eqref{eqn:-Vira Schur} holds for $m=1$.
\vspace{10pt}

In fact, this statement follows from step 1 since equation \eqref{eqn:VirS1} is a trivial equation with both sides equal to $0$
when $m=1$.

\vspace{10pt}
{\bf Step 3}: Equation \eqref{eqn:VirS1} holds for all $m \geq 1$.
\vspace{10pt}

We prove this statement by induction on $m$.
For $m=1$, \eqref{eqn:VirS1} holds trivially since $L_{-1}^S \cdot 1 =0$.
By equation \eqref{eqn:multiSchur},
\[
2 L_{-2}^S \cdot 1 =  p_1^2 \cdot 1=  p_1 s_{(1)} = s_{(2)}+ s_{(1,1)}.
\]
So equation \eqref{eqn:VirS1} also holds for $m = 2$.

Now we consider the case for $m \geq 3$. By the Virasoro bracket relation (see, for example, Section 2.3 in \cite{KR}), we have
\begin{eqnarray*}
(-m+2) L_{-m}^S &=& L_{-(m-1)}^S L_{-1}^S - L_{-1}^S L_{-(m-1)}^S.
\end{eqnarray*}
Since $L_{-1}^S \cdot 1 = 0$, by the induction hypothesis, we have
\begin{equation}
(-m+2) L_{-m}^S \cdot 1
=  - L_{-1}^S L_{-(m-1)}^S \cdot 1
= - L_{-1}^S \sum_{k=1}^{m-1} (-1)^{m-k} \left(-k+\frac{m}{2} \right) s_{(k,1^{m-1-k})}.
\label{eqn:LmSInd}
\end{equation}
Note that $s_{(\cdots, 1, 2, \cdots)}= 0$ by Example \ref{exa:relation Schur}.
By equation \eqref{eqn:-Vira Schur} for $m=1$, which was proved in step 2, we have
\[ L_{-1}^S \cdot s_{(k,1^{r})} = k s_{(k+1, 1^r)} - (r+1) s_{(k, 1^{r+1})}
\]
for all $r \geq 0$. Hence, by equation \eqref{eqn:LmSInd}, we have
\begin{eqnarray*}
(-m+2) L_{-m}^S \cdot 1
&=&    \sum_{k=1}^{m-1} (-1)^{m-k+1} \left(-k+\frac{m}{2} \right)
    \left\{ k s_{(k+1,1^{m-1-k})} - (m-k)  s_{(k,1^{m-k})} \right\} \\
&=&  \sum_{k=1}^m (-1)^{m-k} (-m+2) \left(k- \frac{m+1}{2} \right) s_{(k, 1^{m-k})}.
\end{eqnarray*}
Divided both sides by $(-m+2)$, we obtain the desired formula. This proves the statement in step 3.

The theorem then follows from combination of step 1 and step 3.
$\Box$

\begin{rem}
Equation \eqref{eqn:VirS1} is equivalent to the following identity
\[ \sum_{k=1}^{m-1} p_k p_{m-k} = \sum_{k=1}^m (-1)^{m-k} (2k-m-1) s_{(k, 1^{m-k})} \]
for all $m \geq 1$.
\end{rem}


\vspace{30pt} \noindent
Xiaobo Liu \\
School of Mathematical Sciences \& \\
Beijing International Center for Mathematical Research, \\
Peking University, Beijing, China. \\
Email: {\it xbliu@math.pku.edu.cn}
\ \\ \ \\
Chenglang Yang \\
Beijing International Center for Mathematical Research, \\
Peking University, Beijing, China. \\
Email: {\it yangcl@pku.edu.cn}

\begin{thebibliography}{399}
	

\bibitem[A20]{Alex20} A. Alexandrov,
	{\it Intersection numbers on $\overline{\cal M}_{g, n}$ and BKP hierarchy},
	arXiv:2012.07573.
	
\bibitem[A21]{Alex21} A. Alexandrov,
	{\it Generalized Br\'{e}zin-Gross-Witten tau function as a hypergeometric solution of the BKP hierarchy},
	arXiv:2103.17117.

\bibitem[ASY1]{ASY} K. Aokage, E. Shinkawa and H.-F. Yamada,
{\it Pfaffian identities and Virasoro operators},
Lett. Math. Phys. vol. 110 (2020), p1381-1389.	

\bibitem[ASY2]{ASY2} K. Aokage, E. Shinkawa and H.-F. Yamada,
{\it Virasoro action on the Q-functions},
arXiv:2106.04773.

\bibitem[B]{B} T. H. Baker,
{\it Symmetric function products and plethysms and the boson-fermion correspondence},
J. Phys. A: Math. Gen. 28(1995) 589-606.	

\bibitem[CK]{CK} D. Cox and S. Katz,
{\it Mirror Symmetry and Algebraic Geometry}, Amer. Math. Soc., Providence, RI, 1999.

\bibitem[DJM]{DJM} E. Date, M. Jimbo, and T. Miwa,
{\it Solitons: Differential Equations, Symmetries and Infinite Dimensional Algebras},
Cambridge University Press, (2000).


\bibitem[EHX]{EHX} T. Eguchi, K. Hori, and C. Xiong,
{\it Quantum Cohomology and Virasoro Algebra}, Phys. Lett. B 402 (1997), 71-80.


\bibitem[J]{J}N. Jing, {\it Vertex Operators and Hall-Littlewood Symmetric Functions},
Adv. Math., 87, 226-248.

\bibitem[JL]{JL}N. Jing and N. Liu, {\it The Green Polynomials via Vertex Operators},
arXiv: 2104.04411.

\bibitem[KR]{KR}V. Kac and A.Raina,
{\it Highest Weight Representations on Infinite Dimensional Lie Algebras},
Word Scientific Publishing, 1987

\bibitem[L]{L} D. E. Littlewood, {\it On Certain Symmetric Functions},
Proc. London Math. Soc. (3) 11 (1961), 485-498.



\bibitem[LY1]{LY} X. Liu and C. Yang,
{\it Schur Q-Polynomials and Kontsevich-Witten Tau Function}, arXiv:2103.14318.

\bibitem[LY2]{LY2} X. Liu and C. Yang,
{\it Q-Polynomial expansion for Brezin-Gross-Witten Tau-Function}, arXiv:2104.01357.

\bibitem[Mac]{Mac} I. G. Macdonald, {\it Symmetric Functions and Hall Polynomials},
Clarendon Press, Oxford, (1995).

\bibitem[MMMR]{MMMR} A. Mironov, V. Mishnyakov, A. Morozov, and R. Rashkov,
	{\it Virasoro versus superintegrability, Gaussian Hermitian model}, arXiv:2104.11550.

\bibitem[MM20]{MM20} A. Mironov and A. Morozov,
	{\it Superintegrability of Kontsevich matrix model}, arXiv:2011.12917.

\bibitem[MM21]{MM21} A. Mironov and A. Morozov, {\it Generalized Q-functions for GKM},
arXiv: 2101.08759.

\bibitem[Mor]{Mor} A. O. Morris, {\it A Note on the Multiplication of Hall Functions},
J. London Math. Soc. 39 (1964), 481-488.

\bibitem[Mor76]{Mor76} A. O. Morris, {\it A Survey on Hall-Littlewood Functions and their Applications to Representing Theory},
Lecture Notes in Math., Vol. 579, Springer, Berlin(1977), 136-154.




\bibitem[Y]{Y} Y. You,
    {\it Polynomial solutions of the BKP hierarchy and projective representations of symmetric groups,}
    Infinite-Dimensional Lie Algebras and Groups (Luminy-Marseille, 1988), Adv. Ser. Math. Phys 7 (1989), 449-464.


\end{thebibliography}
\end{document}